# Region-of-interest micro-focus CT based on an all-optical inverse Compton scattering source

Yue Ma[a,b], Jianfei Hua[a,*], Dexiang Liu[a], Yunxiao He[a,b], Tianliang Zhang[a], Jiucheng Chen[a], Fan Yang[a], Xiaonan Ning[a], Zhongshan Yang[a], Jie Zhang[a,b], Chih-Hao Pai[a], Yuqiu Gu[b], and Wei Lu[a,*]

[a] Department of Engineering Physics, Tsinghua University, Beijing 100084, China
[b] Science and Technology on Plasma Physics Laboratory, Laser Fusion Research Center, China Academy of Engineering Physics, Mianyang, Sichuan 621900, China

Micro-focus computed tomography (CT), enabling the reconstruction of hyperfine structure within objects, is a powerful nondestructive testing tool in many fields. Current X-ray sources for micro-focus CT are typically limited by their relatively low photon energy and low flux. An all-optical inverse Compton scattering source (AOCS) based on laser wakefield accelerator (LWFA) can generate intense quasi-monoenergetic X/gamma-ray pulses in the keV-MeV range with micron-level source size, and its potential application for micro-focus CT has become very attractive in recent years due to the fast pace progress made in LWFA. Here we report the first experimental demonstration of high-fidelity micro-focus CT using AOCS (~70 keV) by imaging and reconstructing a test object with complex inner structures. A region-of-interest (ROI) CT method is adopted to utilize the relatively small field-of-view (FOV) of AOCS to obtain high-resolution reconstruction. This demonstration of the ROI micro-focus CT based on AOCS is a key step for its application in the field of hyperfine nondestructive testing.

Inverse Compton scattering (ICS) based on the colliding of an energetic electron beam with an intense laser pulse can generate high-flux, energy-tunable quasi-monoenergetic short X/gamma-ray pulses (1), with potential applications in biology, material science and nuclear physics (2-8). ICS based on conventional accelerator has been developed since 1990s (9-11), but its further applications are quite limited due to its relatively large scale and cost. With the intensive development of LWFA in the past decade (12-15), electron beams with hundreds of MeV energy can be easily obtained within an acceleration distance of just a few millimeters, using a table-top scale ultra-short high-power laser. By colliding this electron beam with another intense laser pulse or simply with the reflected laser pulse using a plasma mirror, energy-tunable (keV to MeV level), high-flux ($\gtrsim 10^8$ photons per pulse) ultra-short AOCS photon pulse can be readily generated (8, 16-23). Due to the tiny spot size of the electron beam at the exit of LWFA, the AOCS typically has a micron-level source size (16), which is especially preferable for high-resolution imaging application, such as micro-focus CT for precise nondestructive testing (3-5).

To date, only AOCS-based projection imaging has been demonstrated (16, 17). Here we report a successful AOCS-based micro-focus CT experiment by imaging and reconstructing a test object using a ~70 keV AOCS driven by a 10 TW Ti:sapphire laser (24). AOCS typically has a limited FOV due to its relatively small cone-angle proportional to $1/\gamma$ ($\gamma$ is the Lorenz factor of the electron beam) (1), therefore an ROI CT scheme is adopted in the experiment. In this scheme, only the volume of interest within the sample (fine inner structures of the test object) is scanned to obtain high-resolution reconstruction. Indeed, high-fidelity reconstruction algorithms of ROI CT have been well developed in the past decade (25-29), and it is especially advantageous for small cone-angle scanning of refined inner structures within a large object.

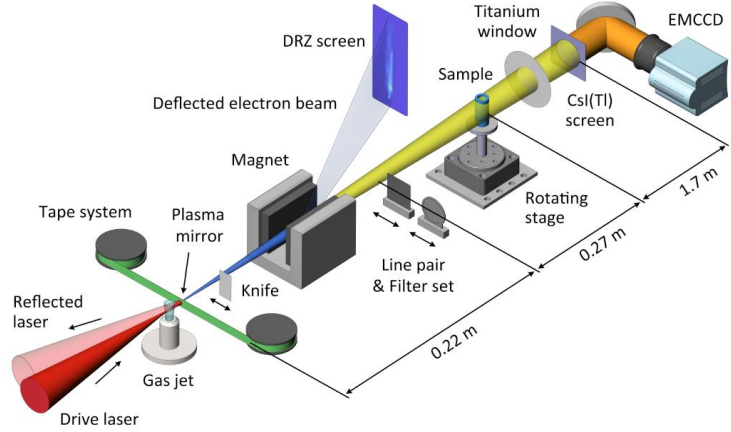

Fig. 1. Schematic layout of the experiment. The AOCS X-ray pulse (yellow) is generated by colliding a LWFA electron beam (blue) with the reflected laser pulse (red) using a plasma mirror. When X-ray photons hit the CsI(Tl) scintillation screen, visible light flashes (orange) are triggered and then captured by an EMCCD.

## Results

**AOCS generation and characterization.** In the experiment, the AOCS X-ray pulses are generated with a scheme using a plasma mirror (16), as illustrated in Fig. 1. The electron beams (~76 MeV, ~120 pC) are produced through LWFA by focusing an intense laser pulse (40 fs, 10 TW) near the front edge of a 2 mm gas jet with mixed gases (99.5% He + 0.5% $N_2$). In Fig. 2, the average energy spectrum and angular divergence of the electron beams for 80 consecutive shots are shown, with an average energy spread of 70% (FWHM) and average divergences of 15/12 mrad (horizontal/vertical). The drive laser pulse existing the gas jet immediately gets reflected by a plasma mirror formed when the laser hits a 100 μm polyethylene terephthalate (PET) tape mounted on an automatic stepping system. AOCS X-ray photons are then generated when the electron beam collides with the reflected laser pulse. In Fig. 3 A, the X-ray intensity profile measured using a CsI(Tl) scintillation screen coupled with an EMCCD is shown, with an average photon yield of $3 \times 10^7$ per shot and divergences of 33/24 mrad (horizontal/vertical).

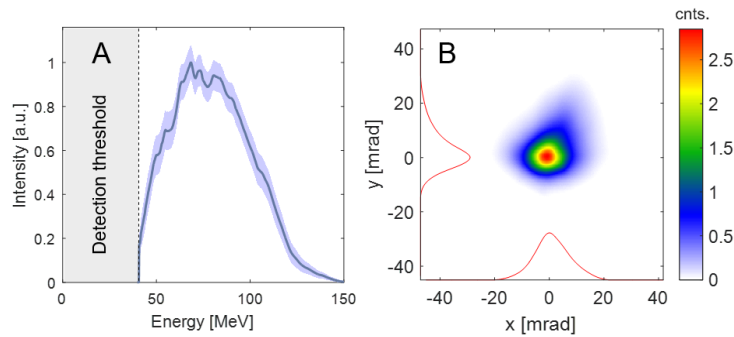

Fig. 2. LWFA electron beam characterization. A. The average spectrum of the electron beams for 80 consecutive shots measured by an electron spectrometer based on a dipole magnet, where the blurred curve shows its FWHM error and the detection threshold is marked by a black dashed line. B. The average angular divergence of the electron beams for 80 consecutive shots.

The X-ray intensity attenuation profile through a multi-sector filter set is measured to reconstruct the spectrum of AOCS using an Expectation Maximization (EM) algorithm (30, 31), as shown in Fig. 3 B and C. The measured on-axis X-ray spectrum has an average energy of ~71keV, in reasonable agreement with a numerical estimation (~66 keV) based on 3D PIC simulation using the experiment parameters (*materials and methods*). The broad bandwidth of the spectrum (~122 keV) is mainly caused by the relatively wide energy spread (70%) of the electron beams and the enlarged bandwidth of the colliding laser induced by the red shift in the plasma (32-34).

To characterize the source size of AOCS, a knife-edge technique (16) is utilized in the experiment. A 300-μm-thick copper plate with a sharp edge is inserted in the X-ray path 1.3 cm downstream of the AOCS, which gives a geometric magnification of ~170 on the detector. In Fig. 3 D, a single-shot image of the sharp edge and its line-out are shown, together with three line-outs from numerical diffraction simulation of the sharp-edge with three different X-ray source sizes (0 μm, 1 μm and 2 μm). Near the knife edge portion, the experimental line-out matches with the simulation line-out very well for 1 μm source size, suggesting the single-shot source size of AOCS is about 1 μm. The accumulated source size for 80 consecutive shots is also measured with the same method to be ~14 μm, as shown in Fig. 3 E. The enlargement of the source size is mainly due to the combined effects of pointing jitters of the laser (~4 μm) and the electron beam (~10 μm caused by the injection process) (31), and both jitters can be further reduced through optimization.

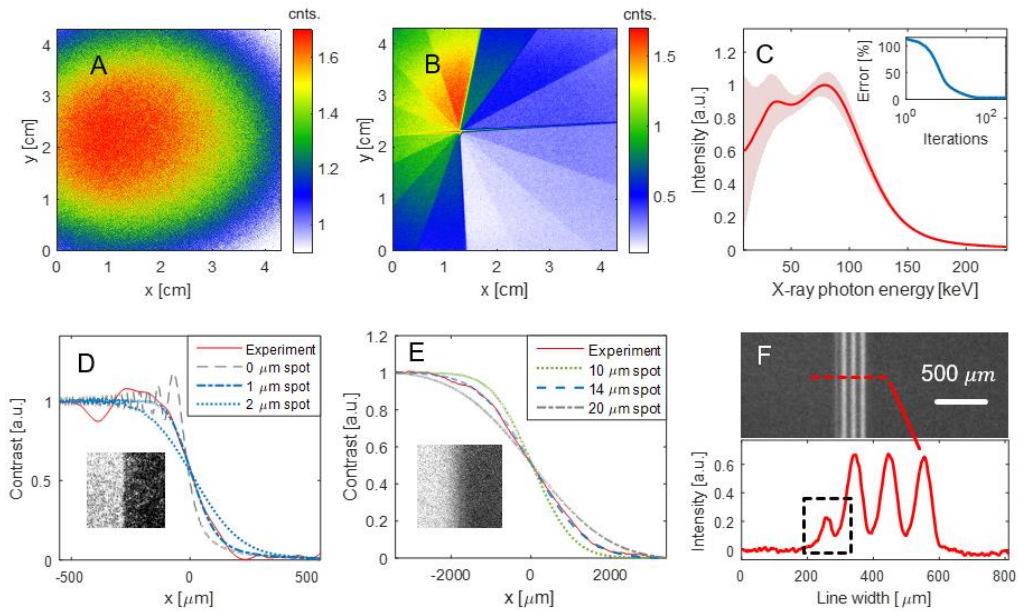

Fig. 3. AOCS characterization. A. An accumulated X-ray profile on the detector for 50 consecutive shots. B. The X-ray intensity profile attenuated by a multi-sector filter set (composed of 7 aluminum plates with the thickness from 0.35 mm to 8 mm and 7 copper plates with the thickness from 0.35 mm to 9 mm) for 50 consecutive shots. C. The on-axis X-ray spectrum deduced using the data in B, where the blurred curve shows its error bar, and the inset shows the calculation error after each iteration. The single-shot source size (D) and the accumulated source size for 80 consecutive shots (E) are measured by the knife-edge technique, and the insets are the images of the sharp edge. F. The accumulated projection image of a 10 lp/mm line pair for 80 consecutive shots and a line-out of it, where a ~20 μm defect within the black dashed square is clearly identified.

Sharp projection image of a 10 lp/mm line pair (Fig. 3 F) is also taken with a geometric magnification of ~10, which confirms that all three lines have a contrast of 70%. Furthermore, a ~20 μm fine defect can also be clearly detected with a contrast of 38%, suggesting the imaging resolution should be better than 20 μm. Indeed, a theoretical estimation of the imaging resolution based on the source size, the detector resolution and the geometric magnification can be deduced to be 13.6 μm (*materials and methods*), in reasonable agreement with the experimental measurement.

**CT imaging and reconstruction.** To demonstrate the AOCS-based ROI micro-focus CT, a test object (Fig. 4 A)

composed of a 1.5 mm thick aluminum tube and a set of 130 μm thick hollow copper pins of a LEMO connector is imaged. The ROI is chosen as the top portion of the pins (the hollow curved structures within the red dashed circle). The object is placed 0.49 m downstream of the AOCS, with a geometric magnification of 4.5 on the detector. In Fig. 4 B, a 60-shots accumulated projection image of the object is shown, where the ROI in the image is highlighted by a red dashed rectangle. The X-ray transmission rate within the ROI varies from 20% to 60% with a local signal-to-noise ratio (SNR) in the range of 10 to 17, sufficient for high-quality reconstruction (2). To estimate the imaging resolution, the Modulation Transfer Function (MTF) of a line-out across the aluminum tube edge (black dashed line in Fig. 4 B) is plotted in Fig. 4 C, suggesting a ~16 μm (~31 lp/mm) resolution with 10% contrast, in good agreement with the theoretical estimation (15.6 μm).

The CT scanning is then performed in 31 directions over 180 degrees, with each projection obtained by 60 shots accumulation. The ROI of the object is reconstructed from only the corresponding portion of the projection images (within the red dashed rectangle in Fig. 4 B) using a compressed sensing (CS) algorithm (26-28). This algorithm has good tolerance to the local SNR, and can generate precise reconstruction with relatively few projections (35-37). A 3D reconstruction of the ROI is shown in Fig. 4 D, and one slice of tomographic image is illustrated in Fig. 4 E. The curved hollow structures of the copper pins are clearly evident in both images. In addition, the average attenuation coefficient of the copper pins can also be calculated to be 8.2 $cm^{-1}$ from the reconstruction data, which suggests an average photon energy about 74 keV, in consistence with the measured X-ray spectrum.

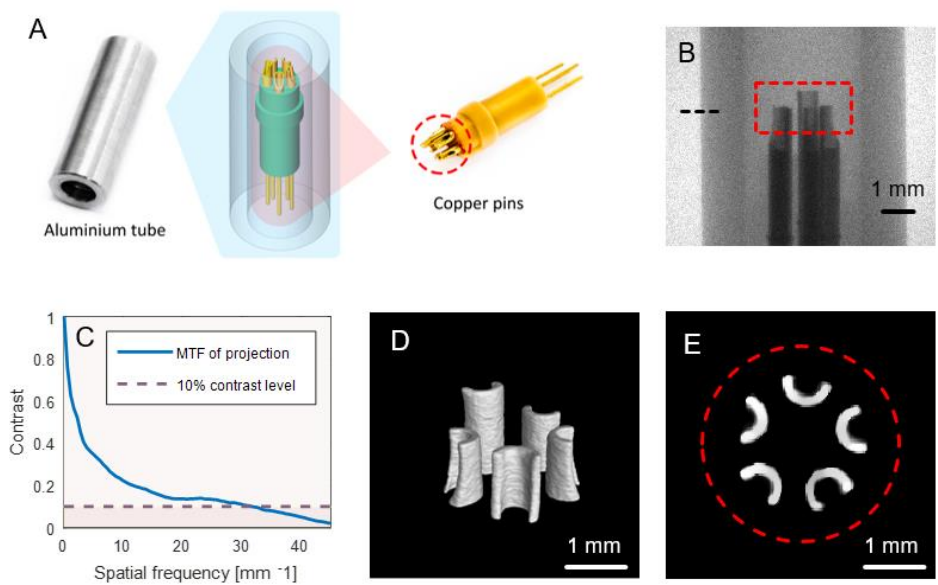

Fig. 4. AOCS-based ROI micro-focus CT imaging. A. An illustration of the test object, with the ROI highlighted by a red dashed circle. B. 60-shots accumulated projection image. C. MTF of the tube edge marked by a black dashed line in the projection image B. D. The 3D reconstructed image within the ROI. E. One slice of the tomographic image in D. The ROI in B and E are highlighted by a red dashed rectangle and circle, respectively.

**Discussion:** AOCS can also generate high-energy (up to 10s MeV) gamma-ray pulses with high flux and micron level source size on a table top (21, 38), and these features make it a unique source for micro-focus CT to image large objects made of dense materials. Current micro-focus CT powered by X-ray tube typically has low flux broadband photons with energy well below MeV (7). On the other hand, for the widely used MeV bremsstrahlung gamma-ray source based on RF accelerators, the CT imaging resolution is quite limited due to its millimeter-level source size (39, 40). Therefore, a high-energy micro-focus CT based on AOCS could be a very attractive solution for high-resolution imaging of dense objects like turbine blades in aircraft engines (typically made of nickel based alloy) (41). Here a numerical CT simulation for such an object using AOCS-based gamma-ray is presented to illustrate its potential.

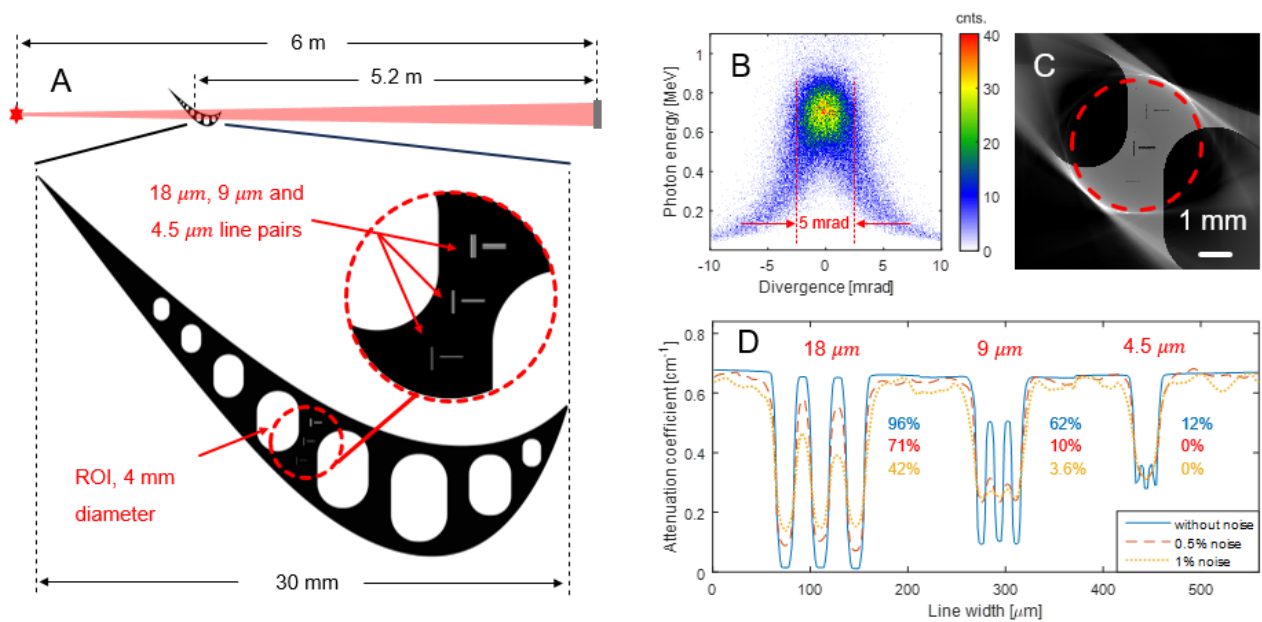

Fig. 5. Numerical simulation of AOCS-based ROI micro-focus CT for a turbine blade. A. The simulation layout: a turbine blade is placed in the X-ray path, with three sets of line pairs embedded in the ROI (within the red dashed circle). B. Photon energy and divergence distribution of the AOCS (divergence within 5 mrad is used in the CT scanning). C. The reconstructed ROI of the turbine blade. D. Line-outs of the reconstructed line pairs with different noise level (contrasts of the line pairs are labeled).

In this simulation, the ROI of the blade (within the red circle in Fig. 5 A) is scanned by a 0.7-MeV AOCS with 4-μm (FWHM) source size in 500 directions over 360 degrees, and then reconstructed using the same algorithm adopted in this experiments (*materials and methods*). In Fig. 5 C, the reconstruction of the ROI is shown, where the fine features (three sets of line pairs) on the blade can be clearly identified. In Fig. 5 D, the line-outs of these three line pairs are also shown to have sufficiently good contrasts (96% for 18 μm width, 62% for 9 μm width and 12% for 4.5 μm width), suggesting an imaging resolution about 4.5 μm. This challenging application is sensitive to the noise of the projections, thus reconstructions of the blade with different projection noise (0.5% and 1%) are performed, and the corresponding line-outs are also plotted in Fig. 5 D, indicating that a noise level below 0.5% is required to achieve a sub-10 μm resolution. Due to the high photon flux of AOCS, such level of projection noise can be easily obtained using a commercial X-ray detector.

To summarize, an experiment of AOCS-based ROI micro-focus CT has been carried out, proving its feasibility on high-resolution reconstruction of complex structures. A numerical simulation to image a dense object (turbine blade) is also performed, showing AOCS's potential application at high energy. With further improvements on source characteristics (source size, spectrum and photon flux) and the detection system (collection efficiency and spatial resolution), compact AOCS-based ROI micro-focus CT may become an important new tool for hyperfine nondestructive testing in the near future.

**Materials and Methods**
**Laser system.** The experiments are performed using a 10 TW Ti:sapphire laser system in Tsinghua University (24).

Laser pulses with 450 mJ energy and 40 fs (FWHM) duration are focused by an f/18 off-axis parabolic mirror to a spot of 9 μm (FWHM diameter) with 60% energy enclosed. In the focused region, the laser intensity is $I = 8.8 \times 10^{18} W \cdot cm^{-2}$, and the corresponding normalized vector potential is $a_0 = 2$.

**X-ray detector.** The X-ray detector is composed of a CsI(Tl) scintillation screen (Hamamatsu J8734), an optical transmission system and an EMCCD (ANDOR DU888E). When the X-ray pulses hit the scintillation screen, visible light photons are emitted, and ~0.2% of them are focused on EMCCD by the optical transmission system. The detector has an effective view of about $4.3 * 4.3$ cm$^2$, and its resolution is ~50 μm.

**Imaging Resolution.** The spatial resolution of projection imaging is typically determined by the X-ray source size **a**, the pixel size of the detector **d** and the geometric magnification **M** as follow (42):

$$R = \frac{\sqrt{d^2 + [a(M-1)]^2}}{M} .$$

For CT imaging, the resolution of the reconstructed tomographic image is also related to the number of the projections according to the Nyquist sampling theorem (43), and the best CT imaging resolution (equal to that of each projection image) can be obtained by providing sufficient number of projections (44, 45). In principle, the CS-based ROI CT algorithm can enable accurate reconstruction using relatively small number of projections (35-37). In our experimental case, 31 projections are indeed sufficient for a high-quality reconstruction.

**Estimation of the X-ray spectrum.** For a nearly head-to-head colliding of the electron beams with the laser pulses in the experiment, the on-axis average energy of AOCS X-ray photons can be estimated as $\omega_s = 4\gamma^2 \omega_r/(1 + a_r^2/2)$ (1), where $\gamma$ is experimentally measured to be ~150, $\omega_r$ and $a_r$ are the angular frequency and the normalized vector potential of the reflected laser pulse respectively. To get an estimation of the X-ray spectrum, $\omega_r$ and $a_r$ are deduced from 3D particle-in-cell (PIC) simulation with OSIRIS (46) using parameters close to the experimental conditions. At the exit of the 2-mm plasma (density of $1.8 \times 10^{19} cm^{-3}$), simulation gives a red-shifted average $\omega_r \approx 0.89 \cdot \omega_0$ ($\hbar\omega_0 = 1.55$ eV) (32-34) and an $a_r \approx 1.3$ (assuming a reflectivity of 90% by the plasma mirror), suggesting an on-axis average photon energy of ~66 keV, in reasonable agreement with experimental measurements.

**CT imaging simulation.** The numerical simulation of high-energy AOCS-based ROI micro-focus CT to image a turbine blade is performed using a MATLAB code. A schematic layout of the simulation is illustrated in Fig. 5 (a), where a nickel-based turbine blade with three sets of line pairs (width of 4.5 μm, 9 μm and 18 μm) inside the ROI is located 0.8 m downstream of the AOCS, with a geometric magnification of 7.5 on the detector (pixel size of 30 μm). The AOCS adopted in the simulation is pre-generated using a Monte-Carlo (MC) code CAIN (47), by colliding a 190 MeV electron beam (12% FWHM energy spread, 1 μm FWHM spot size) with a 0.6 J laser pulse (50 fs FWHM pulse duration, 40 μm waist radius). The obtained AOCS beam has a source size of 1 μm (FWHM) and an on-axis average energy of 0.7 MeV with 38% bandwidth, as shown in Fig. 5 (b). The accumulated source size of AOCS is assumed to be 4 μm by choosing a source position jitter of 3 μm.

**Acknowledgments:** The authors thank Prof. Liang Li and Mr. Siyuan Zhang for a fruitful discussion of the CT algorithm. The authors also thank Mr. Huien Huo, Mr. Junjiang Li and Dr. Yuchi Wu for the support of X-ray detection. This work was supported by the National Natural Science Foundation of China Grants No. 11535006, No. 11991071, No. 11775125, No. 11875175 and Tsinghua University Initiative Scientific Research Program.

**Author Contributions:** W.L. and J.H. conceived the project. Y.M., J.H., D.L., Y.H., T.Z., J.C., F.Y., X.N, Z.Y, J.Z., C.H.P. and

* **Corresponding authors:** Jianfei Hua (jfhua@tsinghua.edu.cn), Wei Lu (weilu@tsinghua.edu.cn)

1. Labate L, Tomassini P, & Gizzi LA (2017) Inverse Compton Scattering X-ray Sources. *Handbook of X-ray Imaging: Physics Technology*:309-323.
2. Cole JM*, et al.* (2018) High-resolution μCT of a mouse embryo using a compact laser-driven X-ray betatron source. *Proceedings of the National Academy of Sciences* 115(25):6335-6340.
3. Holdsworth DW & Thornton MM (2002) Micro-CT in small animal and specimen imaging. *Trends in Biotechnology* 20(8):S34-S39.
4. Golab A, Ward CR, Permana A, Lennox P, & Botha P (2013) High-resolution three-dimensional imaging of coal using microfocus X-ray computed tomography, with special reference to modes of mineral occurrence. *International journal of coal geology* 113:97-108.
5. Badea C, Drangova M, Holdsworth DW, & Johnson G (2008) In vivo small-animal imaging using micro-CT and digital subtraction angiography. *Physics in Medicine Biology* 53(19):R319.
6. Ho ST & Hutmacher DW (2006) A comparison of micro CT with other techniques used in the characterization of scaffolds. *Biomaterials* 27(8):1362-1376.
7. Hale R*, et al.* (2015) High-resolution computed tomography reconstructions of invertebrate burrow systems. *Scientific data* 2:150052.
8. Chen S*, et al.* (2013) MeV-energy X rays from inverse Compton scattering with laser-wakefield accelerated electrons. *Physical review letters* 110(15):155003.
9. Sprangle P, Ting A, Esarey E, & Fisher A (1992) Tunable, short pulse hard x‐rays from a compact laser synchrotron source. *Journal of applied physics* 72(11):5032-5038.
10. Kim K-J, Chattopadhyay S, & Shank C (1994) Generation of femtosecond X-rays by 90 Thomson scattering. *Nuclear Instruments Methods in Physics Research Section A: Accelerators, ectrometers, Detectors Associated Equipment* 341(1-3):351-354.
11. Lee K, Cha Y, Shin M, Kim B, & Kim D (2003) Relativistic nonlinear Thomson scattering as attosecond x-ray source. *Physical Review E* 67(2):026502.
12. Tajima T & Dawson JM (1979) Laser electron accelerator. *Physical Review Letters* 43(4):267.
13. Lu W*, et al.* (2007) Generating multi-GeV electron bunches using single stage laser wakefield acceleration in a 3D nonlinear regime. *Physical Review Special Topics-Accelerators and Beams* 10(6):061301.
14. Blumenfeld I*, et al.* (2007) Energy doubling of 42 GeV electrons in a metre-scale plasma wakefield accelerator. *Nature* 445(7129):741.
15. Brunetti E*, et al.* (2010) Low emittance, high brilliance relativistic electron beams from a laser-plasma accelerator. *Physical review letters* 105(21):215007.
16. Phuoc KT*, et al.* (2012) All-optical Compton gamma-ray source. *Nature Photonics* 6(5):308.
17. Döpp A*, et al.* (2016) An all-optical Compton source for single-exposure x-ray imaging. *Plasma Physics and Controlled Fusion* 58(3):034005.
18. Schwoerer H, Liesfeld B, Schlenvoigt HP, Amthor KU, & Sauerbrey R (2006) Thomson-backscattered x rays from laser-accelerated electrons. *Physical review letters* 96(1):014802.
19. Powers ND*, et al.* (2014) Quasi-monoenergetic and tunable X-rays from a laser-driven Compton light source. *Nature Photonics* 8(1):28.

20. Chen S, *et al.* (2016) Shielded radiography with a laser-driven MeV-energy X-ray source. *Nuclear Instruments and Methods in Physics Research Section B: Beam Interactions with Materials and Atoms* 366:217-223.
21. Liu C, *et al.* (2014) Generation of 9 MeV γ-rays by all-laser-driven Compton scattering with second-harmonic laser light. *Optics letters* 39(14):4132-4135.
22. Tsai H-E, *et al.* (2015) Compact tunable Compton x-ray source from laser-plasma accelerator and plasma mirror. *Physics of Plasmas* 22(2):023106.
23. Zhu C, *et al.* (2018) Inverse Compton scattering x-ray source from laser electron accelerator in pure nitrogen with 15 TW laser pulses. *Plasma Physics Controlled Fusion* 61(2):024001.
24. Jian-Fei H, *et al.* (2015) Generating 10–40 MeV high quality monoenergetic electron beams using a 5 TW 60 fs laser at Tsinghua University. *Chinese Physics C* 39(1):017001.
25. Li L, Kang K, Chen Z, Zhang L, & Xing Y (2009) A general region-of-interest image reconstruction approach with truncated Hilbert transform. *Journal of X-ray Science Technology* 17(2):135-152.
26. Yu H & Wang G (2009) Compressed sensing based interior tomography. *Physics in medicine biology* 54(9):2791.
27. Yang J, Yu H, Jiang M, & Wang G (2010) High-order total variation minimization for interior tomography. *Inverse problems* 26(3):035013.
28. Ueda R, Nemoto T, & Kudo H (2017) Practical interior tomography with small region piecewise model prior. *Medical Imaging 2017: Physics of Medical Imaging*, (International Society for Optics and Photonics), p 101320P.
29. Han Y, Gu J, & Ye JC (2017) Deep learning interior tomography for region-of-interest reconstruction. *arXiv preprint arXiv:.10248*.
30. Lange K & Carson R (1984) EM reconstruction algorithms for emission and transmission tomography. *Comput Assist Tomogr* 8(2):306-316.
31. Guo B, *et al.* (2019) High-resolution phase-contrast imaging of biological specimens using a stable betatron X-ray source in the multiple-exposure mode. *Scientific reports* 9(1):7796.
32. Tsung F, Ren C, Silva L, Mori W, & Katsouleas T (2002) Generation of ultra-intense single-cycle laser pulses by using photon deceleration. *Proceedings of the National Academy of Sciences* 99(1):29-32.
33. Pai C-H, *et al.* (2010) Generation of intense ultrashort midinfrared pulses by laser-plasma interaction in the bubble regime. *Physical Review A* 82(6):063804.
34. Nie Z, *et al.* (2018) Relativistic single-cycle tunable infrared pulses generated from a tailored plasma density structure. *Nature Photonics* 12(8):489.
35. Candès EJ, Romberg J, & Tao T (2006) Robust uncertainty principles: Exact signal reconstruction from highly incomplete frequency information. *IEEE Transactions on information theory* 52(2):489-509.
36. Sidky EY, Kao CM, & Pan X (2006) Accurate image reconstruction from few-views and limited-angle data in divergent-beam CT. *Journal of X-ray Science and Technology* 14(2):119-139.
37. Chang M, *et al.* (2013) A few-view reweighted sparsity hunting (FRESH) method for CT image reconstruction. *Journal of X-ray Science Technology* 21(2):161-176.
38. Yan W, *et al.* (2017) High-order multiphoton Thomson scattering. *Nature Photonics* 11(8):514.
39. Izumi S, Kamata S, Satoh K, & Miyai H (1993) High energy X-ray computed tomography for industrial applications. *IEEE transactions on nuclear science* 40(2):158-161.
40. De Chiffre L, Carmignato S, Kruth J-P, Schmitt R, & Weckenmann A (2014) Industrial applications of computed tomography. *CIRP annals* 63(2):655-677.
41. Lakshmi M, Mondal A, Jadhav C, Dutta B, & Sreedhar S (2013) Overview of NDT methods applied on an aero engine turbine rotor blade. *Insight-Non-Destructive Testing Condition Monitoring* 55(9):482-486.
42. Hussein AE, *et al.* (2019) Laser-wakefield accelerators for high-resolution X-ray imaging of complex microstructures. *Scientific reports* 9(1):3249.

43. Zhao Z, Gang G, & Siewerdsen J (2014) Noise, sampling, and the number of projections in cone‐beam CT with a flat‐panel detector. *Medical physics* 41(6Part1):061909.
44. Bracewell RN & Riddle A (1967) Inversion of fan-beam scans in radio astronomy. *The Astrophysical Journal* 150:427.
45. Gordon R, Bender R, & Herman GT (1970) Algebraic reconstruction techniques (ART) for three-dimensional electron microscopy and X-ray photography. *Journal of theoretical Biology* 29(3):471-481.
46. Fonseca RA*, et al.* (2002) OSIRIS: A three-dimensional, fully relativistic particle in cell code for modeling plasma based accelerators. *International Conference on Computational Science*, (Springer), pp 342-351.
47. Chen P, Horton-Smith G, Ohgaki T, Weidemann A, & Yokoya K (1995) CAIN: Conglomerat d'ABEL et d'Interactions Non-lineaires. *Nuclear Instruments Methods in Physics Research Section A: Accelerators, pectrometers, Detectors Associated Equipment* 355(1):107-110.